\documentclass[twoside,english]{article}
\usepackage{charter}
\usepackage[T1]{fontenc}
\usepackage[latin9]{inputenc}
\usepackage{geometry}
\usepackage{color}
\usepackage{amsthm}
\usepackage{amsmath}
\usepackage{amssymb}
\usepackage{graphicx}
\usepackage{esint}

\makeatletter
\@ifundefined{definecolor}
 {\usepackage{color}}{}
\@ifundefined{definecolor}{\usepackage{color}}{}
\usepackage{latexsym}\usepackage{graphics}\@ifundefined{definecolor}{\@ifundefined{definecolor}{\@ifundefined{definecolor}
 {\usepackage{color}}{}
}{}}{}

\usepackage{babel}

\usepackage{babel}

\usepackage{babel}

\makeatother

\usepackage{babel}
\begin{document}

\title{A Mean-Field Theory for Coarsening Faceted Surfaces}

\author{Scott A. Norris and Stephen J. Watson}

\date{\today}
\maketitle
\begin{abstract}
A mean-field theory is developed for the scale-invariant length distributions
observed during the coarsening of generic one-dimensional faceted
surfaces. This theory closely follows the LSW theory of Ostwald ripening
in two-phase systems \cite{lifshitz-slezov-1959:lsw-theory,lifshitz-slyozov-1961:lsw-theory,wagner-1961:lsw-theory},
but the mechanism of coarsening in faceted surfaces requires the addition
of convolution terms recalling the work of Smoluchowski \cite{smoluchowski-1916:coagulation-1}
and Schumann \cite{schumann-1940:continuous-coagulation} on coalescence,
and results in a novel coupling between the convolution and the rate
of facet loss. As a generic framework, the theory concisely illustrates
how the universal processes of facet removal and neighbor merger are
moderated by the system-specific mean-field velocity describing average
rates of length change. For a simple, example facet dynamics associated
with the directional solidification of a binary alloy, agreement between
predicted scaling state and that observed after direct numerical simulation
of 40,000,000 facets is reasonable given the limiting assumption of
non-correlation between neighbors; relaxing this assumption is a clear
path forward toward improved quantitative agreement with data in the
future.
\end{abstract}

\section{Introduction}

In many examples of faceted surface evolution, a \emph{facet velocity
law} giving the normal velocity of each facet can be observed, assumed,
or derived. Examples of such dynamic laws describe growth of polycrystalline
diamond films from the vapor \cite{kolmogorov-1940:geometric-selection,drift-1967:diamond-evolution-selection},
evolution of faceted boundaries between two elastic solids \cite{gurtin-voorhees-1998},
the evaporation/condensation mechanism of thermal annealing \cite{watson-2006:prl},
and various solidification systems \cite{pfeiffer-1985:faceted-silicon,shangguan-hunt-1989:binary-solidification-facet-dynamics,emmot-bray-1996,watson-2003:physicaD,norris-davis-2007:dssa}.
Such velocity laws are typically configurational, depending on surface
properties of the facet such as area, perimeter, orientation, or position,
and reduce the computational complexity of evolving a continuous surface
to the level of a finite-dimensional system of ordinary differential
equations. This theoretical simplification enables and invites large
numerical simulations for the study of statistical behavior. This
has been done frequently for one-dimensional surfaces \cite{pfeiffer-1985:faceted-silicon,shangguan-hunt-1989:binary-solidification-facet-dynamics,emmot-bray-1996,watson-2003:physicaD,norris-davis-2007:dssa,wild-1990:texture-diamond,thijssen-1992:diamond-dynamic-scaling,paritosh-srolovitz-1999:faceted-films,zhang-adams-2002:FACET,zhang-adams-2004:FACET-followup},
while less frequently for two-dimensional surfaces due to the necessity
of handling complicated topological events \cite{watson-2006:prl,thijssen-1995:diamond-simulation-2,barrat-1996:3D-CVD-diamond,russo-smereka-2000:level-set-facets,smereka-2005:level-set-facets,norris-watson-2007:threefold}.
Such inquiries reveal that many of the systems listed above exhibit
coarsening -- the continual vanishing of small facets and the increase
in the average length of those that remain. Notably, these systems
also display \emph{dynamic scaling}, in which common geometric surface
properties approach a constant statistical state, which is preserved
even as the length scale increases.

The dynamic scaling behavior of coarsening faceted surfaces recalls
the process of Ostwald ripening \cite{ostwald-1900:ripening}, in
which small solid-phase grains in a liquid matrix dissolve, while
larger grains accrete the resulting solute and grow in a scale-invariant
way at late times. Indeed, it was observed some time ago that facets
of alternating orientations on a one-dimensional surface are analogous
to alternating phases of a separating two-phase alloy \cite{mullins-1961:linear-facet-growth,cabrera-1963:symposium},
and the Cahn-Hilliard equation \cite{cahn-hilliard-1958:ch-equation}
which models phase separation has been used, in modified form, to
describe several different kinds of faceted surface evolution \cite{liu-metiu-1993:dynamics_of_phase_separation,golovin-1998:convective_cahn_hilliard_model,norris-davis-2007:dssa}.
More distantly related coarsening systems exhibiting dynamic scaling
include coarsening cellular networks describing soap froths and polycrystalline
films \cite{stavans-1993,fradkow-udler-1994,frost-thompson-1996,thompson-2001,schliecker-2002},
and films growing via spiral defect \cite{schulze-kohn-1999:geometric-spiral-growth}.
In all of these cases, the system is characterized by a network of
evolving boundaries which separate domains of possibly differing composition,
and exhibit coarsening and convergence toward scale-invariant steady
states.

Since dynamic scaling pushes complex systems into a state which can
be approximately characterized by just a few statistics, it is natural
to seek simplified models which approximately mimic the resulting
scaling laws and scaling functions. The canonical example of this
approach is the celebrated theory of Lifshitz, Slyozov, and Wagner
describing Ostwald ripening \cite{lifshitz-slezov-1959:lsw-theory,lifshitz-slyozov-1961:lsw-theory,wagner-1961:lsw-theory}.
Generically, such an approach selects a distribution of some quantity,
and includes just enough of the total system behavior to specify the
effective behavior of that quantity -- for example, the original LSW
theory first identifies the average behavior of particles as a function
of size, and uses that result to identify a continuity equation describing
distribution evolution. Ideas of this kind have been applied to several
of the higher-order cellular systems introduced above -- for soap
froths \cite{beenakker-1986:bubble-mean-field,flyvbjerg-1993:random-neighbor-model},
polycrystals \cite{barmak-etal-2011-DCDSA}, and spiral-growth films
\cite{schulze-kohn-1999:geometric-spiral-growth}. To the extent that
such approaches mirror experimental data, they can yield valuable
physical insight which cannot be gained by considering single particles,
nor even by direct numerical simulation of larger ensembles. However,
to date no similar attempt has been made for evolving faceted interfaces,
which is somewhat surprising given the wide variety of examples of
purely faceted motion, and past success in applying mean-field analyses
to coarsening these systems. 

In this work, therefore, we take a first step in that direction by
introducing a framework for describing the distribution of facet lengths
in 1D faceted surface evolution. Our approach closely resembles the
LSW theory of Ostwald ripening, in that a number density $n(l,t)$,
of facets of length $l$ at time $t$, is transported by a known length-dependent
effective velocity law. However, whereas vanishing drops in Ostwald
ripening simply exit the system, each vanishing facet on a coarsening
surface  causes its two immediate neighbors to join together. Accounting
for this process of merging requires the consideration of a convolution
integral reminiscent of equations due to Smoluchowski \cite{smoluchowski-1916:coagulation-1}
and Schumann \cite{schumann-1940:continuous-coagulation} describing
coagulation (see also Menon \cite{menon-etal-2010-TAMS}). But since
the removal rate of small facets from the system necessarily sets
the rate of the concomitant merger of those neighboring facets, a
particularly novel non-local coupling arises wherein the probability-flux
at the origin is found to weight the convolution integral. We therefore
arrive at the resulting novel evolution equation for the probability
distribution $\rho\left(l,t\right)$ of facets of length $l$ at time
$t$ to be 
\begin{equation}
\frac{\partial\rho}{\partial t}+\frac{\partial}{\partial l}\left[\rho v\right]=-\rho\left(0,t\right)v\left(0,t\right)\int_{0}^{l}\rho(l-s,t)\rho(s,t)\,\text{d}s,\label{eqn: main-result}
\end{equation}
where the \emph{velocity}, $v(l,t)$, takes the special form 
\begin{equation}
v(l,t)=\widehat{\mathcal{V}}\left[l,L(t)\right]\quad\text{with}\quad L(t):=\int_{0}^{\infty}l\rho(l,t)\text{d}l
\end{equation}
for a derived and prescribed \emph{mean-field velocity} law, $\widehat{\mathcal{V}}(l,L)$,
which encodes \emph{an effective rate-of-change of length} for facets
of length $l$ purely in terms of that length, $l$, and the mean
facet length $L$. The left hand side of Equation \ref{eqn: main-result},
and in particular the appearance there of the mean-field velocity
$\widehat{\mathcal{V}}(l,L)$, is precisely where our theory mimics
the essential transport concept underlying the LSW theory. In contrast,
the convolution on the right hand side of \ref{eqn: main-result}
is strongly reminiscent of coagulation models Last, we note that the
time-dependent function multiplying the convolution, $\left(\rho*\rho\right)$,
\[
\mathcal{R}=-\rho\left(0,t\right)v\left(0,t\right),
\]
is a rate of probability flux at the origin, which encodes the link
between facets shrinking to zero and the concomitant merger of the
two neighboring facets.

\section{An Example Problem: From Modeling to Morphometrics\label{sec: example-dynamics}}

\subsection{An Example Coarsening Dynamical System}

\begin{figure}
\begin{centering}
\includegraphics{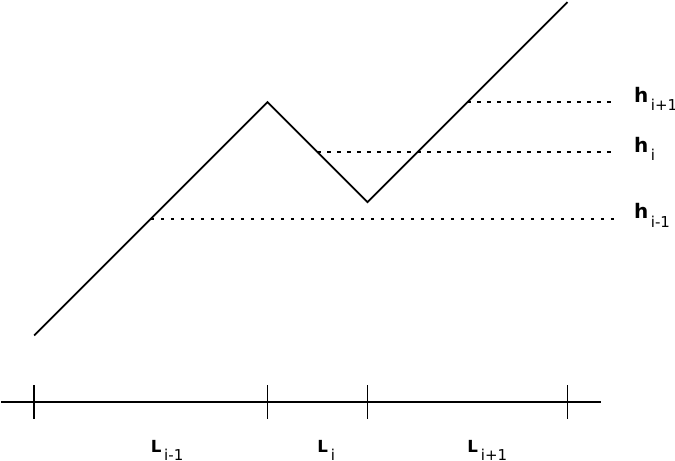} 
\par\end{centering}

\caption{A diagram illustrating a representative facet and its two neighbors.
Here $h_{i}$ denotes the \emph{mean height} of the center facet,
while $L_{i}$ denotes its width (length). Similarly for the neighbors
of the center facet.}

\label{fig: representative-facet-1D} 
\end{figure}
During the directional solidification of a strongly anisotropic binary
alloy, small-wavelength faceted surfaces develop. If the alloy is
solidified above a critical velocity, a layer of supercooled liquid
is created at the interface, which drives a coarsening instability
governed by the facet dynamics \cite{norris-davis-2007:dssa} 
\begin{equation}
\mathcal{V}_{i}=\cos\left(\omega\right)\, h_{i}\,,\label{eqn: ds-facet-dynamics}
\end{equation}
where $\mathcal{V}_{i}$, $(-1)^{i}\omega$ and $h_{i}$ are the (instantaneous)
normal velocity, fixed facet angle (the \textbf{Wulff angle}) and
mean height of the $ith$ facet respectively. Letting $l_{i}(t)$
denote the length of the $ith$ facet at time $t$, we will now show
that, between coarsening events, one can derive from Equation \ref{eqn: ds-facet-dynamics}
a dynamical system for these facet lengths. First, we note the general,
\emph{kinematic} relation 
\begin{equation}
\frac{\text{d}l_{i}}{\text{d}t}=\frac{(-1)^{i}}{\sin\left(2\omega\right)}\left(\mathcal{V}_{i+1}-\mathcal{V}_{i-1}\right);\label{eqn: kinematic-relation}
\end{equation}
the prefactor $(-1)^{i}$ reflects our convention that odd (even)
facets have negative (positive) slopes. If we now use the facet velocity
law \eqref{eqn: ds-facet-dynamics} to specify the \emph{dynamics},
we can convert this equation, via elementary geometry (see Figure
\ref{fig: representative-facet-1D}), to a dynamical system on facet
lengths alone: 
\begin{align}
\frac{dl_{i}}{dt} & =\frac{1}{4}\left(2l_{i}-l_{i+1}-l_{i-1}]\right).\label{eqn: length-dsf}
\end{align}
 Thus, even though the normal velocity of a facet depends on its mean
height and slope, the rate of change of its length does not.

Now the morphometric structure of the evolving coarsening faceted
surface is encoded in an ordered facet-length ensemble, denoted $\mathcal{L}[t]:\mathbb{Z}\to(0,\infty)$,
associated with the evolving interface; $\mathcal{L}[t](i):=l_{i}(t)$.
Since the dynamical system \eqref{eqn: length-dsf} results in lengths
which tend to zero, resulting in the merging of the two neighboring
facets (a coarsening event), we need to \emph{re-write} $\mathcal{L}[t]$
at such critical times. To see how, we again consider Figure~\ref{fig: representative-facet-1D},
and imagine that the $jth$ facet $F_{j}$ shrinks to length $0$
at the critical time $t^{*}$. When this happens, this $F_{j}$ obviously
vanishes; but in addition, the two neighbors of $F_{j}$, namely $F_{j-1}$
and $F_{j+1}$ also vanish as independent entities, to be replaced
by a new facet with length equal to the sum of the lengths of its
ancestors. A natural re-indexing of the resultant faceted surface
provides the following \emph{update rule} $\mathcal{L}[t_{*}^{\,+}]$
\begin{equation}
\mathcal{L}[t_{*}^{\,+}](i):=\begin{cases}
\mathcal{L}[t_{*}^{\,-}](i+1)\hfill i\le j-2,\\
\mathcal{L}[t_{*}^{\,-}](j-1)+\mathcal{L}[t_{*}^{\,-}](j+1)\hspace{3em}\hfill i=j,\\
\mathcal{L}[t_{*}^{\,-}](i-1)\quad\hfill i\ge j+2.
\end{cases}\label{eqn: coarsening-rule}
\end{equation}
Taken together, \ref{eqn: length-dsf} and \ref{eqn: coarsening-rule}
constitute a \emph{Coarsening Dynamical System} (CDS), which is the
object of our further study.

\subsection{Numerical Simulation and Morphological Statistics}

For three different initial length distributions, random faceted surfaces
containing 1,000,000 facets were constructed by generating two sets
of numbers obeying that distribution, and scaled to have equal sums;
these random lengths were then interleaved to generate a periodic
faceted surface. The initial empirical distributions of facet lengths
$\varrho\left(l\right)$ were 

\begin{equation}
\begin{aligned}\varrho_{0,\mathrm{comp.}}\left(l\right) & =\frac{10}{18}\chi_{(\frac{1}{10},\frac{19}{10})}\\
\varrho_{0,\mathrm{exp.}}\left(l\right) & =\exp\left(-l\right)\\
\varrho_{0,\mathrm{poly.}}\left(l\right) & =\frac{2}{\left(1+l\right)^{3}}
\end{aligned}
.\label{eqn: initial-condition-variety}
\end{equation}
Hence, we have explored initial distributions of facet lengths with
compact support, exponential decay, and polynomial decay. 

\begin{figure}[t]
\begin{centering}
\includegraphics{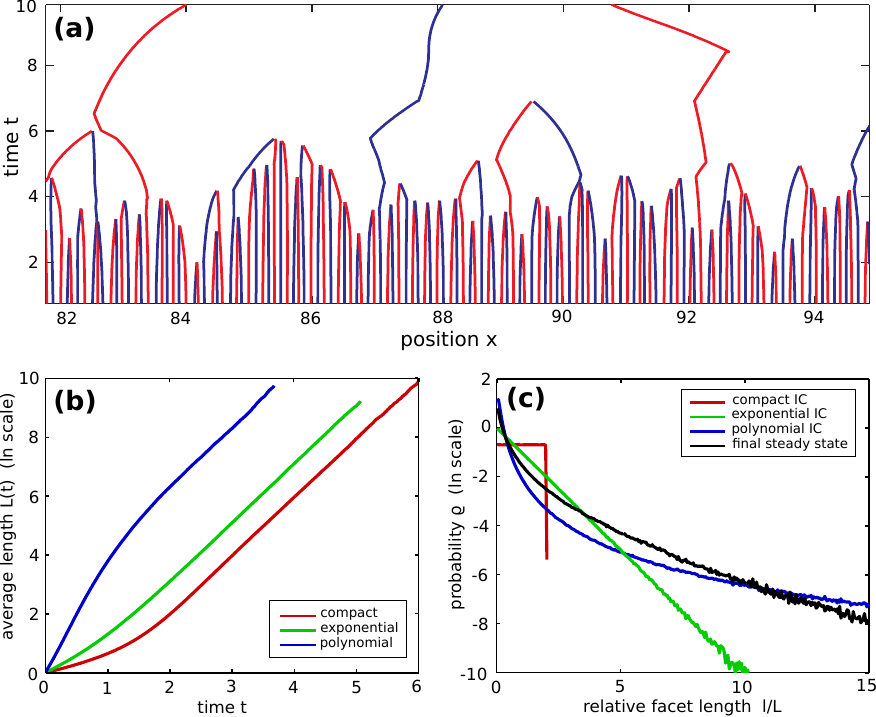}\caption{Survey of coarsening behavior. (a) A representative example of the
evolution of corners between facets (red/blue correspond to hills/valleys).
(b) Facet length scale growth with time for each of the initial conditions
\eqref{eqn: initial-condition-variety} (natural log scale). (c) Numerically
observed scaling function $\mathfrak{P\left(\frac{l}{\left\langle \mathcal{L}\left(t\right)\right\rangle }\right)}$,
compared against each of the initial conditions \eqref{eqn: initial-condition-variety}
(natural log scale).}

\par\end{centering}

\label{fig: coarsening-data} 
\end{figure}

Figure~\ref{fig: coarsening-data}a depicts a representative local
patch of surface evolving under the CDS \eqref{eqn: length-dsf},\eqref{eqn: coarsening-rule}.
There, the locations of facet boundaries (corners) are plotted over
time, and we see many instances of the binary coarsening described
by the update rule \ref{eqn: coarsening-rule}. As the system continues
to evolve, all three initial conditions begin to coarsen exponentially,
with the average facet length $L\left(t\right)\propto e^{1.8t}$ at
late times (Figure~\ref{fig: coarsening-data}b). The arrival at
this rate indicates the attainment of the \emph{scaling state}, in
which the associated empirical probability distribution $\varrho(l,t)$,
of facet lengths $l$ at time $t$, is observed to approach a universal
scale-invariant form 
\begin{equation}
\varrho\left(l,t\right)\xrightarrow{t\to+\infty}\frac{1}{\left\langle \mathcal{L}\left(t\right)\right\rangle }\mathfrak{P}\left(\frac{l}{\left\langle \mathcal{L}\left(t\right)\right\rangle }\right).\label{eqn: ensemble-scaling-structure}
\end{equation}
which is illustrated in Figure~\ref{fig: coarsening-data}c.

\textbf{}

\section{A Mean-Field Theory for Binary-Coarsening Faceted Surfaces\label{sec: general-theory}}

We now turn to a theoretical study of the evolving facet ensemble
associated with the CDS. Our focus is the formulation of a general
theory which sheds light on the empirical probability distribution
$\varrho\left(l,t\right)$. Inspired by LSW theory, our aim is to
formulate an approximation for $\varrho$ by rationally constructing
a mass-transport evolution equation for a theoretical probability
function $\rho(l,t)$, supplemented by sinks/sources which simultaneously
and properly account for the appropriate coarsening mechanism of the
original CDS. Our closed theory emerges from a number of mean-field
approximations of the underlying facet ensemble; such approximations
are often referred to as \emph{mean-field hypothesis}.

\subsection{Preliminaries}

We begin by introducing a number of quantities derived from $\rho\left(l,t\right)$
that are specific to coarsening systems. First, as small facets shrink
to zero and are removed from the system, the \emph{average length}
$L\left(t\right)$ of the remaining facets increases. This monotonically-increasing
quantity is simply the first moment of $\rho$: 
\[
L(t):=\int_{0}^{\infty}l\rho(l,t)\,\mathrm{d}l.
\]
Second, we define the \emph{number density} per unit length, $n\left(l,t\right)$,by
\begin{equation}
n(l,t):=\frac{1}{L(t)}\rho(l,t);\label{eqn: population-density}
\end{equation}
Note that the zeroeth moment $N\left(t\right)=\int n\left(l,t\right)\mathrm{d}l$
- which counts the total number of facets - decreases as $L\left(t\right)$
increases. However, its first moment $\int l\, n\left(l,t\right)\mathrm{d}l\equiv1$
is a conserved quantity, reflecting the fact that the conserved quantity
in our coarsening faceted surface is the total interface length. 

Finally, we presume that under a \emph{closure} or \emph{mean-field
hypothesis} of statistically independent neighboring facet lengths,
it is possible to formulate a probabilistic, length-dependent \emph{transport
rate}
\begin{equation}
v\left(l,t\right):=\left\langle \frac{dl}{dt}\right\rangle \left(l,t\right).
\end{equation}
describing the average rate of length change for facets of length
$l$ at time $t$.

\subsection{Derivation of an Evolution Equation}

The existence of $v\left(l,t\right)$ implies a \emph{population flux}
$J\left(l,t\right)$ 
\begin{equation}
J(l,t):=n(l,t)\, v(l,t).\label{eqn: number-flux}
\end{equation}
For a generic \emph{coarsening} system, $v\left(l,t\right)<0$ in
some neighborhood of $l=0$; the rate of facet removal is then given
by the population flux into the origin
\begin{equation}
R\left(t\right):=-J\left(0,t\right)>0.\label{eqn: facet-removal-rate}
\end{equation}
Under the coarsening rule \eqref{eqn: coarsening-rule}, these facets
are removed from the population, as are the two neighboring facets;
the latter are then replaced with a new facet with length equal to
the sum of the neighbor lengths: 
\[
(\cdots,\mathcal{L}^{-},\mathcal{L},\mathcal{L}^{+},\cdots)\xrightarrow{\mathcal{L}\to0^{+}}(\cdots,\mathcal{L}^{-}+\mathcal{L}^{+},\cdots).
\]
According to our closure hypothesis, the two lost neighbors $\mathcal{L}^{-}$
and $\mathcal{L}^{+}$ are each independently distributed according
to the probability distribution $\rho\left(l,t\right)$. Therefore,
the facet $\mathcal{L}^{-}+\mathcal{L}^{+}$ which replaces them,
being the sum of two random variables, satisfies the \emph{joint probability
function} 
\begin{equation}
\left[\rho*\rho\right]\left(l,t\right)=\int_{0}^{l}\rho(s,t)\rho(l-s,t)\,\mathrm{d}s\label{eqn: joint-pdf}
\end{equation}
Hence, the coarsening process, in addition to transporting particles
out of the domain at $l=0$ at a rate of $R$, induces both a loss
$L_{C}$ and a gain $G_{C}$ of particles with length $l$ at rates
\begin{align}
L_{C}(l,t) & =2R\left(t\right)\rho(l,t)\label{eqn: loss-sink}\\
G_{C}\left(l,t\right) & =R\left(t\right)\left[\rho*\rho\right]\left(l,t\right).\label{eqn: gain-source}
\end{align}
Taken together, the population flux \eqref{eqn: number-flux}, the
neighbor-loss rate \eqref{eqn: loss-sink}, and the neighbor-replacement
rate \eqref{eqn: gain-source} imply a balance law on the number density
given by

\begin{equation}
\frac{\partial n}{\partial t}+\frac{\partial}{\partial l}\left(nv\right)=R\left(\rho*\rho-2\rho\right).\label{eqn: number-continuity}
\end{equation}

We now proceed to recast the left-hand side of Eqn. \eqref{eqn: number-continuity}
purely in terms of the probability density $\rho$. To that end, it
will be useful to introduce the \emph{probability flux}, $\mathcal{J}$,
\[
\mathcal{J}\left(l,t\right):=\rho\left(l,t\right)v\left(l,t\right)=L\left(t\right)J\left(l,t\right)
\]
 and the associated probability flux at the origin, (rate of flux
of probability at 0) 
\[
\mathcal{R}(t):=\mathcal{J}(0,t)=L(t)R\left(t\right).
\]
Noting that integrating the number density \eqref{eqn: number-continuity}
with respect to $l$ over the interval $(0,\infty)$ yields a rate
of facet loss $\frac{\mathrm{d}N}{\mathrm{d}t}$ given by 
\begin{equation}
\frac{\text{d}N}{\text{d}t}=\frac{\text{d}\ }{\text{d}t}\left(\frac{1}{L}\right)=-2R=-2\frac{\mathcal{R}}{L},\label{TNE}
\end{equation}
 we obtain 
\begin{equation}
\frac{\partial n}{\partial t}=\frac{\partial}{\partial t}\left(\frac{\rho}{L}\right)=-2\frac{\mathcal{R}}{L}\rho+\frac{1}{L}\frac{\partial\rho}{\partial t}.\label{eqn: dndt-def}
\end{equation}
Inserting \eqref{eqn: dndt-def} into \eqref{eqn: number-continuity},
we conclude that the governing equation for the probability distribution
$\rho(l,t)$ is 
\begin{equation}
\boxed{\rho_{t}+\left(\rho v\right)_{l}=\mathcal{R}\left(\rho*\rho\right)}.\label{eqn: probability-evolution}
\end{equation}
This equation - our main result - is generic to coarsening faceted
surfaces exhibiting binary coarsening, with the effect of the particular
facet dynamics limited to the term $v\left(l,t\right)$.

\subsection{Predictions: Growth Rate and Scaling State}

To investigate the scaling state of Equations \eqref{eqn: probability-evolution}
and \eqref{eqn: mean-field-velocity-DSSA}, we make the \emph{scaling
hypothesis} that 
\begin{equation}
\rho\left(l,t\right)=\frac{1}{L\left(t\right)}\mathcal{P}\left(\frac{l}{L\left(t\right)}\right)\label{eqn: scaling-hypothesis}
\end{equation}
where
\begin{equation}
\int_{0}^{\infty}\mathcal{P}\left(l\right)\mathrm{d}l=1\qquad\mathrm{and}\qquad\int_{0}^{\infty}l\,\mathcal{P}\left(l\right)\mathrm{d}l=1.\label{eqn: unit-moments-1}
\end{equation}
This hypothesis leads to two main results. First, recalling \eqref{TNE}
and \eqref{eqn: facet-removal-rate}, we obtain a differential equation
on $L\left(t\right):$
\[
\frac{\mathrm{d}}{\mathrm{d}t}\left(\frac{1}{L}\right)=-2\left[n\left(0,t\right)v\left(0,t\right)\right]=-\frac{\mathcal{P}\left(0\right)}{L}.
\]
Hence, the theory predicts that the average facet length grows according
to to the exponential relation
\begin{equation}
L\left(t\right)=\exp\left[\mathcal{P}\left(0\right)t\right];\label{eqn: length-vs-time}
\end{equation}
this form is independent of the particular facet dynamics, which serve
only to choose the constant $\mathcal{P}\left(0\right)$. Second,
upon inserting the ansatz \eqref{eqn: scaling-hypothesis} into \eqref{eqn: probability-evolution},
we obtain for the \emph{scaling function} $\mathcal{P}\left(x\right)$
the governing equation
\begin{equation}
\frac{\mathrm{d}}{\mathrm{dx}}\left[v\left(x\right)\mathcal{P}\left(x\right)\right]=\mathcal{P}\left(0\right)\left\{ \int_{0}^{\infty}\mathcal{P}\left(x\right)\mathcal{P}\left(\xi-x\right)\mathrm{d}\xi+2\frac{\mathrm{d}}{\mathrm{d}x}\left(x\mathcal{P}\left(x\right)\right)\right\} ;\label{eqn: scaling-state-PDE-general}
\end{equation}
this integro-differential equation implicitly defines $\mathcal{P}\left(x\right)$,
and can be solved at least numerically. The results \eqref{eqn: length-vs-time}
and \eqref{eqn: scaling-state-PDE-general} are the central predictions
of our theory.

\section{Comparison of Theory vs Data for our Example Dynamics}

We now apply the generic theory of Section \ref{sec: general-theory}
to the specific example dynamics \eqref{eqn: length-dsf} from Section
\ref{sec: example-dynamics}, and compare the theoretical predictions
on $\rho$ with the empirical statistical data $\varrho$.

\paragraph{Application of the general theory}

To generate a theory specific to the dynamics \eqref{eqn: ds-facet-dynamics},
all that is required is to calculate the mean-field velocity $v\left(l,t\right)$
associated with the dynamics \eqref{eqn: length-dsf}, and insert
it into the relevant generic equations of section \ref{sec: general-theory}.
From \eqref{eqn: length-dsf}, the application of our neighbor-independence
hypothesis leads immediately to 
\begin{equation}
v\left(l,t\right)=\frac{1}{2}\left(l-L\left(t\right)\right).\label{eqn: mean-field-velocity-DSSA}
\end{equation}
Inserting \eqref{eqn: mean-field-velocity-DSSA} into the generic
scaling state equation \eqref{eqn: scaling-state-PDE-general} yields
as the definition of the scaling function for this dynamics the equation
\begin{equation}
\frac{\mathrm{d}}{\mathrm{dx}}\left[\left(x-1\right)\mathcal{P}\left(x\right)\right]=2\mathcal{P}\left(0\right)\left\{ \int_{0}^{\infty}\mathcal{P}\left(x\right)\mathcal{P}\left(\xi-x\right)\mathrm{d}\xi+2\frac{\mathrm{d}}{\mathrm{d}x}\left(x\mathcal{P}\left(x\right)\right)\right\} .\label{eqn: scaling-state-PDE-example}
\end{equation}
By inspection, a solution to this equation happens to be 
\begin{equation}
\mathcal{P}\left(x\right)=\exp\left(-x\right),\label{eqn: scaling-function-example}
\end{equation}
and hence, the average facet length is predicted to grow at late times
as 
\begin{equation}
L\left(t\right)=e^{\mathcal{P}\left(0\right)t}=e^{t}.
\end{equation}

\paragraph{Comparison with Data}

\begin{figure}
\begin{centering}
\includegraphics{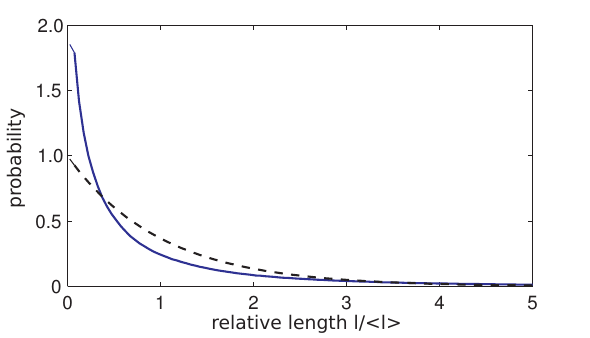} 
\par\end{centering}

\begin{centering}
\caption{Comparison between the theoretically-predicted scaling function $\mathcal{P}\left(x\right)$
(blue) and the empirically-observed function $\mathfrak{P}\left(x\right)$
(black dashes).}

\par\end{centering}

\label{fig: scaling-comparison} 
\end{figure}
We now compare the predictions \eqref{eqn: length-vs-time} and \eqref{eqn: scaling-function-example}
against steady state statistics obtained via direct simulation of
the dynamics \eqref{eqn: ds-facet-dynamics}. First, the observed
statistical coarsening rate of $e^{1.8t}$, although different from
the predicted rate of $e^{t}$, is consistent with the generic prediction
\ref{eqn: length-vs-time}, in that our observed scaling state exhibited
$\mathfrak{P}\left(0\right)\approx1.8$. This is unsurprising, because
the prediction \ref{eqn: length-vs-time} is obtained ultimately from
the conservation of total surface length. The discrepancy between
actual values is due to a difference in the scaling state itself,
shown in Figure \ref{fig: scaling-comparison}. There the predicted
exponential distribution is shown in blue; comparison with the observed
distribution (green) reveals qualitative but not quantitative agreement.
In particular, although both distributions exhibit exponential decay
in the dimensionless relative length $l/L$, the observed distribution
has more of its mass near zero and in the tail, which decays like
$\exp\left(-\frac{x}{2}\right)$ rather than the predicted $\exp\left(-x\right)$.

\paragraph{Diagnostic Tests for the source of disagreement}

\begin{figure}
\begin{centering}
\includegraphics{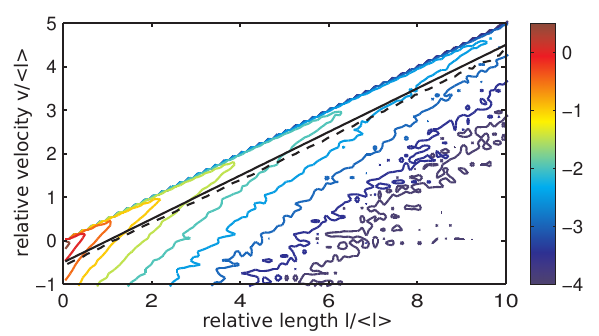} 
\par\end{centering}

\begin{centering}
\includegraphics{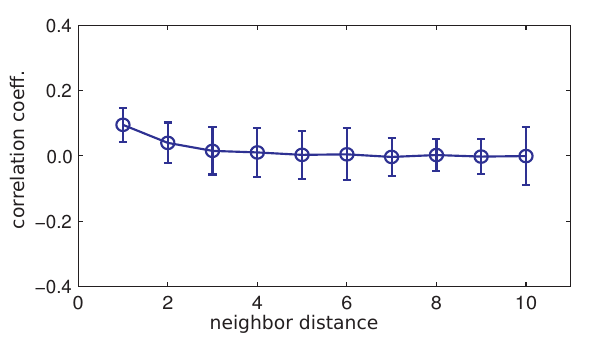}
\par\end{centering}

\begin{centering}
\caption{Diagnostic tests: (a) Contour of the logarithm of the statistical
distribution of dimensionless length/velocity pairs $\varrho(\frac{l}{L},\frac{v}{L})$.
The mean statistical velocity (obtained by integrating in $\frac{v}{L}$)
is plotted as a dotted line, while the predicted velocity $v\left(\frac{l}{L}\right)$
is a solid line. (b) Steady-state correlation of facet lengths as
a function of neighbor distance $d$. Plotted is the correlation coefficient
of all length pairs $\left(l_{i},\, l_{(i+d)\%n}\right)$. }

\par\end{centering}

\label{fig: diagnostics} 
\end{figure}
Seeking the cause of quantitative discrepancies between theory and
experiment, we re-examine the two main assumptions used to construct
the theory: the existence of a statistical mean field velocity, and
the non-correlation of neighboring facet lengths.

\subparagraph{Mean-field velocity.}

To test the validity of the assumption of an effective flux function
$v(l)$, we construct a two-point distribution of length/velocity
pairs $(l,v)$ as they occur in the simulated ensembles. A contour
plot of the logarithm of this distribution is shown in Figure~\ref{fig: diagnostics}.
Finding the mean velocity for each length gives the statistical $v$
(dashes), which turns out to compare favorably with the predicted
$v$ (solid). A slight shift between the curves is observed, which
we attribute to a discretization error associated with binning our
sample data into boxes of width 0.1 in the relative velocity and relative
lengths. So this approximation seems to be reasonable.

\subparagraph{Non-correlation.}

To test the validity of the assumption of non-correlation between
neighboring facets, we measure the correlation coefficient of each
$n^{\mathrm{th}}$-neighbor pair for various $n$. These coefficients
tend toward a steady state as the ensembles evolve, and that state
is shown in figure Figure~\ref{fig: diagnostics}b. There we see
a small but possibly significant positive correlation for at least
the first two neighbors, suggesting that facets of similar size tend
to cluster together - large with large, and small with small. Hence,
the non-correlation assumption fails, though not spectacularly. This
result is not surprising, as the main weakness of the original LSW
theory which inspires this work was also a failure to address correlations;
later generalizations which corrected this deficiency agreed well
with experimental data \cite{marder-1987:ostwald-ripening-correlations}.

\section{Conclusions}

We have presented a mean-field theory for the evolution of length
distributions associated with coarsening faceted surfaces. In the
spirit of LSW theory, a facet-velocity law governing surface evolution
is used to establish a characteristic length-change law; this \emph{mean-field
velocity}, along with a consideration of the effect of binary coarsening
events observed during simulation, leads to a novel continuity equation
governing the evolution of the probability distribution. This equation
combines a transport term similar to LSW theory with a convolution
term reminiscent of coagulation-fragmentation models. However, it
is novel in that the latter process occurs at a rate determined by
the magnitude of transport at the boundary. Our model therefore serves,
apart from the direct application to facet dynamics, as a study of
a new class of mean-field equation.

For a sample facet dynamics associated with binary solidification,
we find the growth rate, scaling state, and coarsening efficiency
predicted by our framework, and then compare the predictions to statistics
obtained from direct numerical simulation of a large facet ensemble.
The results, although not quantitative, agree surprisingly well for
a single-point statistic. In addition, the theory
captures the essential feature of the dynamically scaling state --
a particular mass flux law which drives coarsening by pushing facets
away from the average length, moderated by competing terms describing
coarsening and continuous rescaling, which push mass toward infinity
and zero, respectively. While later improvements to our model addressing
neighbor correlation will undoubtedly increase its predictive capabilities,
these same forces will still balance in the steady state. The model
as presented thus serves as a qualitative explanation of the essential
features of the scaling state, as well as a guide to further research
efforts.

\vspace{0.5in}

\textbf{Acknowledgements.} \,\, SAN was supported by NASA GSRP \#NGT5-50434
for the early stages of this work.\clearpage{}
\bibliographystyle{unsrt}
\bibliography{tagged-bibliography}

\end{document}